\documentclass[aps,prl,twocolumn]{revtex4-1}

\usepackage{amsmath}
\usepackage{amsthm}
\usepackage{amsfonts}

\usepackage{bbm,graphicx,bm,hyperref,color}

\usepackage{lipsum}

\usepackage{multirow}
\usepackage{blkarray, bigstrut}

\def\ave#1{\langle #1 \rangle}
\def\ii{{\rm i}}
\def\sx{\sigma^{\rm x}}
\def\sy{\sigma^{\rm y}}
\def\sz{\sigma^{\rm z}}

\def\1{\mathbbm{1}}

\def\ket#1{{| #1 \rangle}}
\def\bra#1{{\langle #1 |}}

\def\bracket#1#2#3{{\langle #1|#2 | #3 \rangle}}

\def\ee#1{{\rm e}^{#1}}

\def\h{{\mathfrak{D}}}
\def\tauc{{\tau_{\rm c}}}
\def\Uq{{\rm U}_q{\rm (\mathfrak{sl}_2)}}
\def\tSp{{\tilde S}^+}
\def\Jeff{{J_{\rm eff}}}
\def\l{{l}}

\def\tit#1{{\em #1},}

\def\appsym{Methods}
\def\appsu2{Methods}
\def\appnum{Methods}
\def\appcur{Methods}

\newcommand{\new}[1]{{#1}}

\begin{document}

\title{Inhomogeneous SU(2) symmetries in homogeneous integrable U(1) circuits and transport}

\author{Marko \v Znidari\v c}
\affiliation{Physics Department, Faculty of Mathematics and Physics, University of Ljubljana, 1000 Ljubljana, Slovenia}

\date{\today}

\begin{abstract}
Symmetries are important for understanding equilibrium as well as nonequilibrium properties like transport. In translationally invariant extended systems one might expect symmetry generators to also be homogeneous. Studying  qubit circuits with nearest-neighbor U(1) gates we show that this needs not be the case. We find new inhomogeneous screw SU(2) and $\Uq$ symmetries whose generators exhibit a spatial quasi-momentum modulation. They can be viewed as a parameter-dependent generalization of the standard rotational symmetry of the Heisenberg model and can be identified by the Ruelle-Pollicott spectrum of a momentum-resolved propagator. Rich integrability structure is reflected also in transport: picking an arbitrary U(1) gate and varying the gate duration one will transition through different phases, including fractal ballistic transport, Kardar-Parisi-Zhang superdiffusion at the critical manifold that also contains helix states, diffusion, and localization. To correctly explain transport the non-local SU(2) symmetries do not matter, while the inhomogeneous local ones that almost commute with the propagator do.
\end{abstract}





\maketitle

Symmetry is one of the most overarching concepts in physics~\cite{gross}. While in principle just delineating a playing field for dynamics, at low temperatures, for instance, \new{it is} restrictive enough to pin down scenarios to only a handful of possibilities. To classify the phases one essentially needs to know symmetries of the order parameter and the Hamiltonian~\cite{sachdev,wen}. In integrable models the effects of symmetries are the strongest. A symmetry brings with it a conservation law, and integrable systems are, vaguely speaking, systems with an extensive number of conserved quantities~\cite{js,baxter,faddeev}. \new{Symmetry can also influence transport properties of conserved charges, see e.g. Ref.~\cite{enej}.}

How do we find symmetries of a known system? Often this is done by inspection -- in a suitable frame symmetries are obvious. One can also look at the spectrum, namely, a presence of a symmetry will be reflected in spectral multiplets -- a number of eigenstates with the same (quasi)eigenenergy. Studying integrable Floquet quantum circuits with a gate that generalizes the Heisenberg interaction -- a model of direct experimental interest -- we find that all of the above approaches fail. Looking at the Hamiltonian the symmetries will not be clear at all, and the correct multiplets will also be absent.

We show that this is due to a new type of SU(2) and $\Uq$ symmetries whose generators depend on model parameters and are spatially dependent, despite the system being homogeneous -- all two-qubit gates are the same. Furthermore, the symmetry in a finite system will not be exact because the generators commute with the propagator only upto boundary terms. Depending on parameters symmetry generators can be either local, or few-body but non-local. Identifying such unconventional hidden SU(2) symmetry will be crucial to correctly explain the observed bulk properties as quantified by magnetization transport. The non-local SU(2) symmetries do not matter for transport while the local ones do. All will be demonstrated on homogeneous circuits with U(1) preserving gates described by a recently discovered $4$-parameter integrable family~\cite{U1}, a special case being the 2-parameter XXZ gates~\cite{gritsev,vanicat18,ljubotina19,miao24}. Extra parameters will bring new phenomena not found in neither the XXZ circuit nor in the \new{autonomous} Hamiltonian chain. A century since \new{their} discovery~\cite{bethe}, Heisenberg-type integrable models still manage to surprise with beautiful mathematical structures having physical consequences that can be probed in experiments~\cite{jepsen20,google,bloch22,john23,maruyoshi23,ibm,dario,shtanko25}.

\begin{figure}[t!]
  \centerline{\includegraphics[width=3.1in]{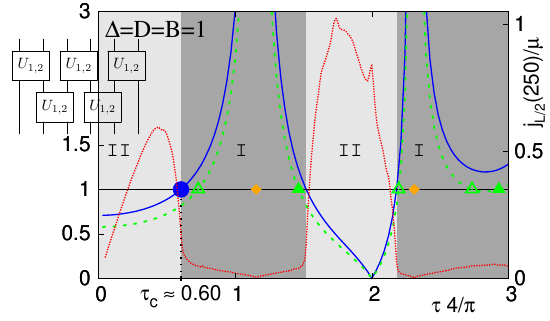}}
  \caption{{\bf Phases as a function of gate duration $\tau$}. When blue curve $|\h|$ (\ref{eq:crit}) crosses $1$ the circuit is critical (\new{e.g.,} blue circle); crossings of green dashed curve (\ref{eq:Uqcond}) mark \new{positions} of non-local SU(2) symmetries (green triangles, Eq.(\ref{eq:Uq2})). Red curve (right axis) \new{is the spin current in the middle of the circuit at $t=250$ starting from a weakly polarized domain wall ($L=1000$),} and indicates the transport type: \new{fractal} ballistic in phase II, diffusive in I, and localization at orange diamonds (Eq.(\ref{eq:loc})). Gate parameters are $D=B=\Delta=1$.}
  \label{fig:phases}
\end{figure}

\begin{flushleft}
  {\em \bf Mischievous symmetries}
\end{flushleft}
\vskip-5mm

Any two-qubit gate with U(1) symmetry, i.e., conserving the magnetization $\sz_1+\sz_2$, can be written as
\begin{eqnarray}
  &U_{1,2}={\rm e}^{-\ii h_{1,2}\tau},\quad\,\, h_{1,2} =\sx_1\sx_2+\sy_1\sy_2+\Delta \sz_1 \sz_2+\notag\\
  &\,\,\,\, +B(\sz_2-\sz_1)+D(\sx_1 \sy_2-\sy_1 \sx_2)+\new{m} (\sz_1+\sz_2).
  \label{eq:H}
\end{eqnarray}
The one-step propagator is denoted by $U$ and is a product of the above gates $U_{k,k+1}$ applied in a brickwall pattern, $U=(\prod_{\l=1} U_{2\l-1,2\l})(\prod_{\l=1} U_{2\l,2\l+1})$, with $L$ qubits (we assume even $L$). It has been shown~\cite{U1} that all such same-gate U(1) circuits with periodic boundary conditions (PBC), regardless of parameter values, are Yang-Baxter integrable. Because the total magnetization $Z=\sum_{\l=1}^L \sz_\l$ is conserved\new{, $m$} under PBC affect only the overall phase and we set it to \new{$m=0$}.

It is known~\cite{ljubotina19} that transport of the XXZ circuit, i.e. at $B=D=0$, varies between ballistic, diffusive, and superdiffusive. Crucial property determining the transport type are symmetries. Especially interesting is superdiffusion in interacting integrable systems~\cite{super,sarang19,dupont20,vir20b,rahul21,ziga20}. It has been explained~\cite{enej} that integrable models with a non-Abelian symmetry, like the SU(2), will generically display superdiffusion (in symmetry-invariant states). To that end we first look at symmetries of our model (\ref{eq:H}). 

Recall that based on the analytical integrability structure \new{of $U$} two phases have been found~\cite{U1}, with \new{the criticality condition} expressed in terms of $\h(\Delta,D,B,\tau)$ as
\begin{equation}
  |\h|=1,\, \h\equiv\frac{\sin{(2\tau\Delta)}}{\sin{(2\tau\Jeff)}}\frac{\Jeff}{\sqrt{1\!+\!D^2}},\, \Jeff\equiv\!\! \sqrt{1\!+\!D^2\!+\!B^2}.
  \label{eq:crit}
\end{equation}
The phase I is obtained for $|\h|>1$, while the phase II for $|\h|<1$. At zero magnetization and infinite temperature transport in the XXZ circuit~\cite{ljubotina19} ($B=D=0$) is diffusive in phase I, ballistic in phase II, with the critical point $|\h|=1$, \new{happening} at $|\Delta|=1$ (in the basic cell $2\Delta\tau,2\tau \in [-\pi/2,\pi/2]$), displaying superdiffusion and Kardar-Parisi-Zhang (KPZ) 2-point correlations~\cite{kpz,evers20}. In the XXZ circuit the critical point therefore coincides with the isotropic \new{XXX} generator where the SU(2) symmetry is obvious. However, for the newly discovered general integrable gate with $B,D\neq 0$ one does not seem to have any obvious SU(2) symmetry at the 3-dimensional critical manifold $|\h|=1$. For instance, setting $D=B=\Delta=1$ the critical condition is satisfied only at special values of $\tau$, the smallest one being $\tauc \approx 0.605535 \frac{\pi}{4}$ (Fig.~\ref{fig:phases}).

\begin{figure}[t!]
  \centerline{\includegraphics[width=2.8in]{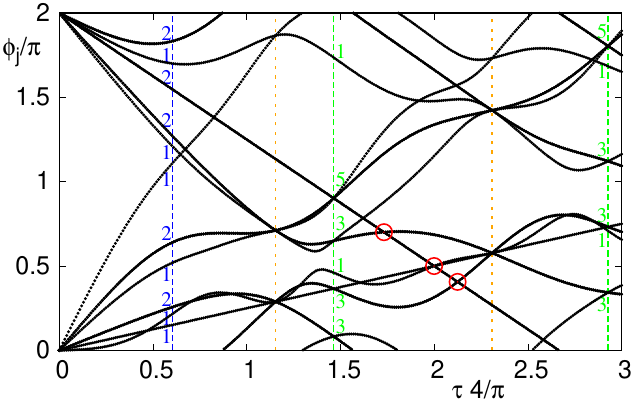}}
  \caption{{\bf Circuit spectrum}. \new{Eigenphases of $U$ are shown as a function of $\tau$ for $L=4$ and $\Delta=D=B=1$.} At the critical point $\tauc$ (blue; circle in Fig.~\ref{fig:phases}) there are no SU(2) multiplets, while at green lines ($\tau_+$ (\ref{eq:Uq2}) for integer $k$) there are. Blue/green numbers are degeneracies. Red circles are extra degeneracies that will be important for fractality. Vertical orange lines are Eq.(\ref{eq:loc}).}
  \label{fig:L4tau}
\end{figure}

To nevertheless identify the presence of any possible non-obvious SU(2) symmetry we have looked at the eigenphases spectrum of $U$. While one could use Bethe ansatz to get the Floquet spectrum~\cite{aleiner21} we simply use numerical diagonalization with a view of possibly using it also on systems with not-yet-known integrability. Namely, if one has an SU(2) symmetry (and no other) one should see corresponding degeneracies: $L$ spins $\frac{1}{2}$ can be coupled into a total spin $s$ running over integer/half-integer values $s=L/2,L/2-1,\ldots$. For instance, \new{two spins can be combined into $s=0$, or $s=1$, i.e., one has one singlet $s=0$ and one triplet $s=1$. For $L=4$ (shown in Fig.~\ref{fig:L4tau}) one can combine the singlet and the triplet of $L=2$ twice into $s=0$, three times into $s=1$, i.e., one has 3 triplets, and once into $s=2$, resulting in the multiplet structure $2[s=0]\oplus 3[1] \oplus [2]$. Because the generators commute with $U$ all $2s+1$ states within a given total spin-$s$ multiplet will have the same eigenvalue. Therefore, for $L=4$ one should see 2 non-degenerate eigenvalues (one for each of $s=0$), 3 multiplets of degeneracy 3, and one 5 times degenerate eigenvalue. For general $L$ and $s$ the number of multiplets of spin $s$ is ${L+1 \choose L/2-s}\frac{1+2s}{L+1}$.}
In Fig.~\ref{fig:L4tau} we show an example of a spectrum as a function of $\tau$ for open boundary conditions (OBC). Surprisingly, there are no multiplets at the critical $\tauc$ (nor for PBC), while on the other hand there are SU(2) multiplets at other non-critical values in phase I. This is puzzling because, as we shall show latter, the transport is superdiffusive at the critical point, while it is diffusive in phase I. It looks as the symmetries are just the opposite of what they should be~\cite{enej}: apparently one has superdiffusion without SU(2), while in the presence of SU(2) one sees diffusion\new{?}

We will resolve this conundrum by: (i) finding a hidden SU(2) symmetry at the critical manifold; the symmetry generators will be spatially dependent even though the propagator $U$ is translationally invariant, (ii) explicitly construct the SU(2) generators, related to $\Uq$ symmetry, at special points in phase I \new{(green lines in Fig.~\ref{fig:L4tau})}, and show that \new{they are non-local}.

\begin{flushleft}
  {\em \bf Inhomogeneous SU(2) at the critical manifold}
\end{flushleft}
\vskip-3mm
For $B=D=0$ one has the well known SU(2) generators, $Z$ and the ladder operators $2S^\pm=\sum_\l \sx_\l \pm \ii\, \sy_\l$. If one has $B=0$ but nonzero $D \neq 0$ things are still simple. Namely, by a unitary rotation $W$~\cite{alcaraz90}
\begin{equation}
  W=\ee{-\ii \vartheta \sum_{\l=1}^L \l \sz_\l},\qquad \tan{(2\vartheta)}=D,
  \label{eq:W}
\end{equation}
one can transform an OBC circuit with $D\neq 0$ to a circuit with $D=0$ (\appsym). The SU(2) generators for $B=0$ are therefore the rotated ones, explicitly $\tSp=\sum_\l \sigma^+_\l \ee{-\ii 2\l\vartheta}$. So-far those generators are exactly the same as for the autonomous XXZ spin chain with the $D$ term. The phase $2\vartheta$ is a nonzero quasi-momentum of the conserved one-site translations operator.

The interesting case is $B\neq 0$. First, we note that a brickwall circuit is invariant under translations by two sites, not by one like the spin chain, and we have to allow for an even/odd site effects. The standard $Z$ still commutes with $U$ so we only have to find the new $\tSp$. The following staggered ansatz will work
\begin{equation}
  \tSp=\sum_\l (\sigma^+_{2\l-1}+\ee{-\ii(2\vartheta-\alpha)}\sigma^+_{2\l})\ee{-\ii 4\l\vartheta}.
  \label{eq:tSp}
\end{equation}
There is a relative phase $\alpha$ \new{between even and odd sites} and a nonzero \new{quasi-}momentum $2\vartheta$ determined by $\tan{(2\vartheta)}=D$. While such $\tSp$ always satisfies SU(2) algebra it does not commute with $U$. In fact, it turns out that regardless of $\alpha$ it never commutes with $U$ (for OBC or PBC) -- this is in accordance with the absence of SU(2) multiplets (Fig.~\ref{fig:L4tau}). However, with an appropriate $\alpha$ such $\tSp$ almost commutes with $U$. That is, in a finite system with OBC one has
\begin{equation}
  U^\dagger \tilde{S}^+ U-\tilde{S}^+ = 0 \,\,+\,\, ({\rm boundary\ terms}),
  \label{eq:comm}
\end{equation}
where boundary terms act nontrivially either on site $1$ or on $L$. Such an almost commutation has been for instance found also for quasi-local charges in the XXZ chain~\cite{prosen11}. Plugging the ansatz for $\tSp$ in Eq.(\ref{eq:comm}) we obtain after some manipulations an explicit expression for the phase,
\begin{equation}
  \ee{\ii\alpha(\tau,\Delta,D,B)}=\h \frac{\cos{(2\tau\Jeff)}}{\cos{(2\tau\Delta)}}-\ii \frac{B}{\sqrt{1+D^2}} \tan{(2\tau\Delta)}.
  \label{eq:alpha}
\end{equation}

Several observations are in place. The resulting SU(2) symmetry holds only at the critical manifold, $|\h|=1$ (where $|\ee{\ii \alpha}|=1$; $\ee{\ii\alpha}=\pm 1$ at $B=0$, which takes care of e.g. $\Delta<0$; \new{alternatively}, $\tan{\alpha}=-\frac{B}{\Jeff}\tan{(2\tau\Jeff)}$). It is exact only in the thermodynamic limit (for two almost commuting \new{Hermitian} operators one can always find two close exactly commuting operators~\cite{hastings}), and is not isotropic. Despite the system being homogeneous\new{, i.e., is translationally invariant under the shift by $2$ sites, the generator is inhomogeneous: on top of the even-odd relative phase $\alpha$ the local generators rotate in the $xy$ plane as one moves along the chain, $\tilde{\sigma}_l^{\rm x} \sim \tilde{\sigma}_l^+ + \tilde{\sigma}_l^- \sim \sx_l \cos{(2l\vartheta)} + \sy_l \sin{(2l\vartheta)}$, where $\tilde{\sigma}^+_{l} \approx \sigma^+_l \exp{(-\ii 2l\vartheta)}$ (that is why we call it a screw symmetry).} Interestingly, the generators explicitly depend on gate parameters. This symmetry is new and is not possible in the Hamiltonian chain $H=\sum_l h_{l,l+1}$ where the $B$ terms mutually cancel.

\begin{figure}[t!]
  \centerline{\includegraphics[width=2.8in]{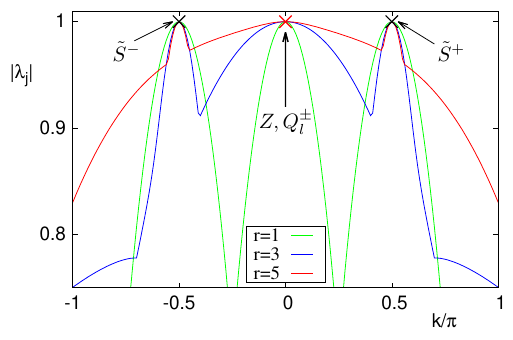}}
  \caption{{\bf Momentum-resolved operator propagator}. \new{The largest Ruelle-Pollicott eigenvalue of $M$ is shown as a function of quasi-momentum $k$} for a critical circuit with $\Delta=D=B=1$ and $\tauc$, \new{and for different truncations $r$.} One can see peaks at $k=\pm \frac{\pi}{2}$ corresponding to screw SU(2) operators $\tilde{S}^\pm$ (\ref{eq:tSp}).}
  \label{fig:rp}
\end{figure}
Considering that such a symmetry is not at all visible in the spectrum, neither for OBC nor for PBC, and that its generators are parameter-dependent and therefore hard to identify by inspection, one might wonder how can one in general find such symmetries? One way is by the recently introduced momentum-resolved operator propagator~\cite{RP24} developed in the context of Ruelle-Pollicott (RP) resonance spectra. Namely, one can write down a linear operator $M$ that propagates operators in an infinite system. Such an operator is unitary, however, if truncated down to \new{translations of} local operators with nontrivial support only on at most $r$ consecutive sites it becomes non-unitary. In a translationally invariant system one can work in a given quasi-momentum block $k$ (for \new{$k=0$} see Ref.~\cite{Prosen}). \new{Such} RP spectra, traditionally used in studies of chaotic systems~\cite{Prosen,RP24}, are useful also in integrable systems~\cite{U1}. Namely, eigenvalues $1$ of $M(k)$ indicate the presence of strictly local conserved operators. Numerically constructing $M(k)$ for our circuit (see Refs.~\cite{RP24,U1} for details) we plot in Fig.~\ref{fig:rp} \new{the largest eigenvalue of $M(k)$ truncated to operators with support on $r=1,3,5$ sites (${\rm dim}[M(k)]=6\cdot 4^{r-1}$).} We can see a degenerate eigenvalue at momentum $k=0$ which is $r$ times degenerate with the corresponding eigenvectors being translationally invariant conserved local charges $Q_p^\pm$~\cite{U1}. In addition though we get two nondegenerate peaks at $k=\pm 4\vartheta$ \new{exactly} corresponding to SU(2) ladder operators (\ref{eq:tSp}). Because they are strictly local and 1-body they are visible already for $r=1$. 

\begin{flushleft}
  {\em \bf Non-local {SU(2)} and $\Uq$ symmetries}
\end{flushleft}
\vskip-3mm
It remains to explain what will turn out to be non-local SU(2) symmetries that do not influence transport (green lines in Fig.~\ref{fig:L4tau}). Quantum group $\Uq$ is important in many areas of mathematics and physics, including integrability due to its deep connection to the $R$ matrix~\cite{kulish81,drinfeld,jimbo}. For $q$ that are not roots of unity ($q^m \neq \pm 1$) the multiplets of $\Uq$ are exactly the same as those of SU(2). Therefore it immediately follows that there also exist generators of SU(2): they can be explicitly constructed for any finite $L$ via diagonalization (\appsu2). We are therefore going to look for $\Uq$ symmetries.

We are inspired by $\Uq$ symmetry observed~\cite{pasquier90,kulish91} in the XXZ chain with OBC and boundary fields given by our $B$ of strength $1+B^2=\Delta^2$ (\new{with} $q+q^{-1}=2\Delta$), for the Floquet setting see Ref.~\cite{miao24}. This can be immediately generalized to $D\neq 0$ using the rotation by $W$ (\ref{eq:W}). A potentially negative sign $s$ of $\Delta$, $s={\rm sign}(\Delta)$, can be flipped by a rotation with $W(\vartheta=\pi/2)$. Provided
\begin{equation}
  \Delta^2=\Jeff^2=1+D^2+B^2,\quad s= {\rm sign}(\Delta),
  \label{eq:Uq1}
\end{equation}
is satisfied, $\Uq$ generators
\begin{eqnarray}
  & S^\pm_q=\!\! \sum_\l q^{-Z_{[1,\l-1]}/2} \otimes {\sigma_\l^\pm}\ee{\mp \ii 2\l(\vartheta+\frac{\pi}{2}\frac{1-s}{2})} \otimes q^{Z_{[\l+1,L]}/2},\nonumber \\
  & \frac{1}{2}(q+q^{-1})=\frac{\Jeff}{\sqrt{1+D^2}},
  \label{eq:genUq}
\end{eqnarray}
where $Z_{[j,p]}\equiv \sum_{\l=j}^p \sz_\l$, commute with OBC $U$ for any $\tau$ (for $s\cdot B>0$ one takes the solution with $q>1$, otherwise $q<1$), as well as with $H=\sum_l h_{l,l+1}$. Together with $Z$ they satisfy the $\Uq$ algebra
\begin{equation}
\!\!\! [Z,S^\pm_q]=\pm 2 S^\pm_q,\,\, [S^+_q,S^-_q]=[Z]_q,\,\, [x]_q\equiv \frac{q^x-q^{-x}}{q-q^{-1}}.
  \label{eq:Uq}
\end{equation}
\begin{figure*}[t!]
  \centerline{
    \includegraphics[width=\textwidth]{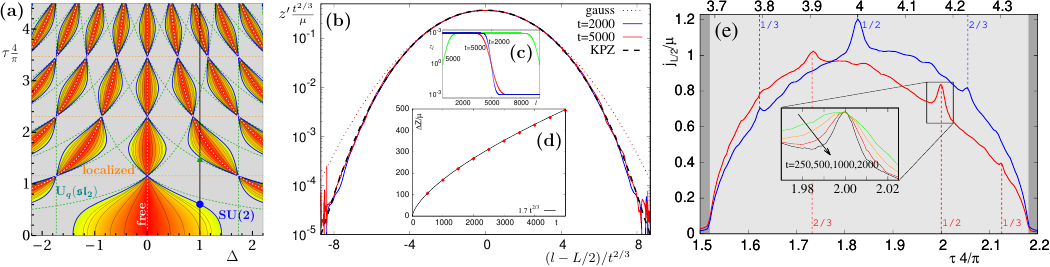}
  }
\caption{{\bf Transport in different phases.} (a) Phase diagram: gray regions are diffusive phase I, colored ballistic phase II (colors at constant $\h$ (\ref{eq:peaks})), blue the critical manifold. Vertical black line is a cross-section shown in Fig.~\ref{fig:phases} and (e), green dashed curves indicate $\Uq$ symmetry, orange ones localization and white dotted curves the non-interacting \new{case}. (b-d) Superdiffusion and KPZ 2-point correlations at the critical $\tauc$ (blue circle in (a)), $L=10^4, \chi=256$: (b) KPZ scaling of the 2-point function, (c) magnetization profiles (red and blue); green points show $\langle \tilde{\sigma}^{\rm x}_l \rangle$ for the initial state polarized in the $\tilde{x}$ direction (see text and \appnum), and (d) transferred magnetization. (e) Ballistic phase and fractal dependence of the current in the middle of the chain at $t=10^3$ on $\tau$, starting from a weakly polarized domain wall ($L=4000$). Red curve (bottom axis) is for the 2nd ballistic region in (a), blue curve (top axis) for the 3rd. Vertical dashed lines and $p/m$ mark strongest peaks (\ref{eq:peaks}). \new{In all plots $\Delta=D=B=1$.}}
  \label{fig:all}
\end{figure*}

The condition (\ref{eq:Uq1}) can be generalized to \new{any $D,B,\Delta$} by realizing that the $\Uq$ is present if
\begin{equation}
  \lvert \sin{(2\tau\Delta)}/\sin{(2\tau\Jeff)} \rvert =1.
  \label{eq:Uqcond}
\end{equation}
This immediately gives \new{two sets} of $\tau$-dependent $\Uq$ points,
\begin{equation}
  \tau_{\pm}=\frac{k\pi}{\Jeff \pm \Delta},
  \label{eq:Uq2}
\end{equation}
with the generator in Eq.~(\ref{eq:genUq}) \new{with $s=-1$ for $\tau_+$ and $s=+1$ for $\tau_-$}. For integer $k$ (full green triangles in Fig.~\ref{fig:phases}) they exactly commute with $U$ for OBC, while for half-integer $k$ (empty green triangles in Fig.~\ref{fig:phases}) they exactly commute under OBC with $\tilde{U}=U\sz_1\sz_L$. 

\new{Now that we have $\Uq$ symmetries we} can construct SU(2) generators (see \appsu2). Expanding them over the basis of Pauli matrices, their locality can be quantified by the range $r$ of a \new{Pauli product} (the largest distance between two non-identity Paulis), and the number $p$ of non-identity Paulis (e.g., two-body next-nearest neighbor terms have $p=2$ and $r=3$). Locality of SU(2) generators is in fact similar to those of $\Uq$ (\ref{eq:genUq}). They are products of $\sigma^\pm_l$ and $\sz_p$, with the highest contribution to $S^+$ (leading order in $q$) coming from $\sigma^+_l$, the next order from $\sigma^+_l \sz_p$, and so on, see \appsu2. The main point relevant for transport is that those generators are not local -- while the weight of many-body terms decays exponentially with $p$ (they are quasi few-body) their range extends over the whole system, $r \sim L$, i.e., terms like $\sigma^+_l \sz_L$ have approximately the same weight as $\sigma^+_l \sz_{l+1}$. It is not clear if such few-body non-local symmetries have physical consequences (sometimes non-local conserved charges do matter~\cite{fagotti,fagotti2}).

For $B=0$ (implying $q=1$) the $\Uq$ condition (\ref{eq:Uqcond}) coincides with the criticality (\ref{eq:crit}) and SU(2) \new{generators} (\ref{eq:tSp}) -- e.g. blue and green curves in Fig.~\ref{fig:all}(a) would overlap. Interestingly, as the field $B$ is switched on the local SU(2) symmetry splits into two symmetries: one acquires a nonzero staggering phase $\alpha$ and stays local -- this is the local SU(2) at the critical manifold; the other keeps a trivial $\alpha$ (translational invariance by one site) but becomes non-local -- this is the $\Uq$. \new{Nonzero $B$ is therefore responsible} for new symmetries not present in neither Floquet nor Hamiltonian XXZ systems.

\begin{flushleft}
  {\em \bf Transport}
\end{flushleft}
\vskip-3mm
Knowing symmetries, in particular the SU(2) one, we are now ready to understand magnetization transport in U(1) integrable circuits at infinite temperature and zero magnetization (Fig.~\ref{fig:all}(a)). At the critical manifold the ``hidden'' inhomogeneous SU(2) symmetry whose generators (\ref{eq:tSp}) are sums of local 1-body terms suggests superdiffusion with a dynamical exponent $z=\frac{3}{2}$, similar as in the standard isotropic XXX circuit~\cite{ljubotina19}. This is indeed what is observed in Fig.~\ref{fig:all}(b). Starting with a mixed weakly polarized domain wall under OBC (initial polarization $\ave{\sz_\l}=\pm \mu=10^{-3}$, see \appnum), we: (i) see a clear superdiffusive growth of the transferred magnetization from the left to the right chain half, $\Delta Z \sim t^{1/z}$, (ii) using the same numerical simulation on $L=10^4$ spins we calculate the infinite temperature correlation function~\cite{kpz} $\langle \sz_0(0) \sz_\l(t) \rangle_{T=\infty} = \lim_{\mu \to 0} \frac{z'}{\mu}$, where $z'\equiv \ave{\sz_{\l-1}(t)}-\ave{\sz_\l(t)}$ (due to slight even-odd staggering we average magnetization over two consecutive sites and calculate the derivative between even sites), showing agreement with the KPZ scaling function $f(\varphi)$~\cite{spohn} (that determines the surface slope correlations in the KPZ equation~\cite{KPZ}) over 4 decades. One could also start with a domain-wall polarized in the $\tilde{x}$ direction, where the conserved magnetization $\sum_l \tilde{\sigma}^{\rm x}_l$, with $\tilde{\sigma}^{\rm x}_l=\cos{\varphi_l} \sx_l+\sin{\varphi_l}\sy_l$, has a quasi-momentum (\ref{eq:tSp}) phase $\varphi_l=2(l+1)\vartheta-(1+\ee{\ii \pi l})\alpha/2$. Interestingly, even starting with a state polarized \new{all} up, $\langle \tilde{\sigma}^{\rm x}_l\rangle=\mu$, i.e., not a domain-wall but a kind of a helix state (see Refs.~\cite{alcaraz90,popkov21,ketterle22,zhang23} for helix states), due to unmatched phases at boundaries a superdiffusive front will spread from the edge (Fig.~\ref{fig:all}(c) and Methods). This explains a mysterious observation \new{(commented on already in Ref.~\cite{kpz} for the standard SU(2))} that in high-precision KPZ simulations a larger $L$ than suggested by only the central superdiffusive lightcone hitting the boundary is needed \new{-- there are in fact two lightcones, one spreading from the center and one spreading from a boundary.}

In phase I, including points with the non-local $\Uq$ symmetry, we find diffusion, see \appnum{ }. Therefore, the non-local nature of SU(2) generators (\appsu2), specifically the long-range 2-body terms $\sigma^+_l\sz_j$ is enough to render such a symmetry irrelevant for transport. On a somewhat \new{related} note we remark that phases brought by strings of $\sz_l$ are enough to break superdiffusion to diffusion even if they act locally, an example being the XX non-local dephasing model~\cite{superXX} that otherwise shows superdiffusion~\cite{yupeng,yupengdis}. Even though based on exact SU(2) multiplets it looked that the connection~\cite{enej} between transport and SU(2) symmetry was violated, everything is fine provided one understands that (i) the symmetries need to hold only in the thermodynamic limit; in finite systems there can be boundary violations, and (ii) the generators need to be \new{(quasi)}local.

In the middle of phase I one also has points with localization because the gate becomes diagonal. This happens when $\h$ is infinite and one resonantly annuls the hopping,
\begin{equation}
2\Jeff \tau=k\pi,\qquad k \in \mathbb{Z}.
\label{eq:loc}
\end{equation}
Those points are visible also in spectra as extra degeneracies (\new{orange lines in} Fig.~\ref{fig:L4tau} ; for even $L$ just $L$ different eigenphases).

Finally, there is the ballistic phase II. In the ballistic phase the speed of transport (i.e., the Drude weight) will have fractal dependence on any generic parameter. For instance, picking an arbitrary fixed set of $\Delta,D,B$, and varying the gate duration $\tau$ one will repeatedly cross through phases II (Fig.~\ref{fig:all}(a)) within which the transport speed is a fractal. This is shown in Fig.~\ref{fig:all}(e) where we simply plot a finite-time proxy for the Drude weight given by the current in the middle of the chain after evolving the initial weakly polarized domain-wall. For definition of the current see \appcur. The fractal dependence comes from quasi-local conserved charges~\cite{prosen13} which can be constructed at roots of unity $q=\ee{\ii \pi p/m}$ using finite-dimensional representations of $\Uq$. In the XXZ spin chain those commensurate conditions are $\Delta=\cos{(\pi p/m)}$ for any co-prime integers $p$ and $m$, while in the XXZ circuit~\cite{ljubotina19} it was identified that one of the $R$-matrix parameters $\eta$ had to be a rational multiple of $\pi$. With that in mind, and the fact that our criticality condition (\ref{eq:crit}) is simple in terms of the $R$-matrix parameters~\cite{U1}, as well as identification of $\Uq$ symmetries in the diffusive phase (\ref{eq:Uq2}), we can generalize the above conditions to
\begin{equation}
  \h=\cos{\left(\pi p/m\right)}.
  \label{eq:peaks}
\end{equation}
$\h$ (\ref{eq:crit}) therefore plays the role of a generalized anisotropy.  The strongest fractal peaks occur for small values of $m$; in Fig.~\ref{fig:all}(e) we indicate location of the ones for $m=2$ and $m=3$, which are also locations of extra degeneracies in the Floquet spectrum (red circles in Fig.~\ref{fig:L4tau}). For finite times fractal peaks are broadened with their width scaling as $\sim 1/\sqrt{t}$, see also Ref.~\cite{ljubotina19}. Interestingly, fractal structure has been observed also in the steady state density of the XXZ chain under appropriate boundary driving~\cite{aash24}. Ballistic transport is especially fast at the non-interacting points $\h=0$, or explicitly $\tau/(\pi/4)=2k/\Delta$ \new{with integer $k$}. Observe that transport at those points is not always the fastest, e.g., in Fig.~\ref{fig:all}(e) the red peak at the free point $p/m=1/2$ is smaller than the one at $2/3$.

\begin{flushleft}
  {\em \bf Discussion}
\end{flushleft}
\vskip-3mm
All properties that we discussed trivially apply to the corresponding Hamiltonian system obtained for $\tau \to 0$, however, the most interesting parameter $B$ is absent as it only produces boundary fields. We provide a generalization of the XXZ criticality condition to any U(1) circuit. Importantly, it shows that the fact that criticality and isotropy coincide in the XXZ model is accidental. Those two are inequivalent notions, and it is the inhomogeneous SU(2) symmetry and not the isotropy that matters. \new{While the propagator $U$ and the conserved charge $Z$ are homogeneous, i.e. from the quasi-momentum sector $k=0$, the symmetry generator that determines transport is inhomogeneous with nonzero $k$.} It would be interesting to better understand how the local SU(2) symmetry of the XXZ model splits into the local staggered SU(2) and the non-local $\Uq$ symmetry \new{upon} nonzero $B$.  

\new{In different phases the model shows different transport type,} ranging from fractal ballistic, to diffusion, and KPZ superdiffusion, as well as localization. Results also highlight a subtle nature of symmetries. To observe superdiffusion SU(2) generators have to be local, though they do not need to be exactly conserved. \new{In accordance with different phase properties we expect that other quantities will also exhibit transitions. An example is the observation of transitions as a function of $\tau$ for the XXZ gate (called Trotter transition)~\cite{vernier23}. Our condition $|\h|=1$ generalizes that result to any U(1) gate.}

Regarding the fractal transport one could repeat steps of Ref.~\cite{ljubotina19} (numerical solution of the Fredholm eq.) to get the fractal structure directly for $t \to \infty$. Systems with an exact quantum group symmetries are of special interest due to their simplified integrability~\cite{nepo18}, an example being $\Uq$ and the XXX spin chain, or the \new{anisotropic} XXZ chain with special boundary fields~\cite{pasquier90}. In our Floquet system it would be interesting to understand if the exact $\Uq$ symmetries of $U$ at special points have any physical consequences, and if with appropriate boundary terms they can be made exact or almost exact, including at other parameters. A direct connection to the underlying integrability and Bethe equations, and if there are other conserved operators with non-zero quasi-momentum, is also unexplored.

\new{Ability to use} any U(1) gate is experimentally attractive -- no fine tuning is required. \new{Moving away from exact integrability the question of stability of observed phenomena is important not just theoretically but also experimentally. To rigorously define transport one needs an infinite system size, however, as can be seen e.g. by red curve in Fig.~\ref{fig:phases} already a finite size and time data on the current in a simple quench experiment is a good indicator. While probing the inhomogeneous nature of symmetry generators directly might be experimentally challenging, a helix state forming at boundaries starting with initially transversally polarized state could be an alternative.} Better understanding such superdiffusive spin helix states is open.

\begin{flushleft}
  {\em \bf \large Methods}
  \end{flushleft}

\begin{flushleft}
  {\bf \new{Generators} of SU(2) from multiplets}
\end{flushleft}
\vskip-3mm
\label{app:su2}

Here we describe how to construct the generators of SU(2) that commute with $U$ provided one has a system with the same multiplets as SU(2) and knows the $Z$ generator. In our brickwall circuits the $Z$ \new{is} conserved by construction, and we know that the system has exactly the same SU(2) multiplets for all parameters with $\Uq$ symmetry and non root of unity $q$.

We shall construct the raising operator $S^+$ by first diagonalizing $U$ and identifying degenerate blocks. In each degenerate block corresponding to one SU(2) multiplet and spanned by $\{ \ket{\xi_l} \}$ we find a basis of $Z$ by diagonalizing the projection $\bracket{\xi_l}{Z}{\xi_{l'}}$, obtaining the eigenstates $\ket{m}$ of $\frac{1}{2}Z$ in the block. The operator $S^+$ in the block with spin $s$ is now by definition equal to $\sum_{m=-s}^s \sqrt{s(s+1)-m(m+1)}\ket{m+1}\bra{m}$, while the whole operator $S^+$ is a direct sum of such terms over all spin $s$ multiplets. Once we have $S^+$, the lowering operator is $S^-=(S^+)^\dagger$, thereby obtaining $X$ and $Y$ that have SU(2) algebra and commute with $U$ by construction.

\begin{figure}[t]
  \centerline{\includegraphics[width=2.8in]{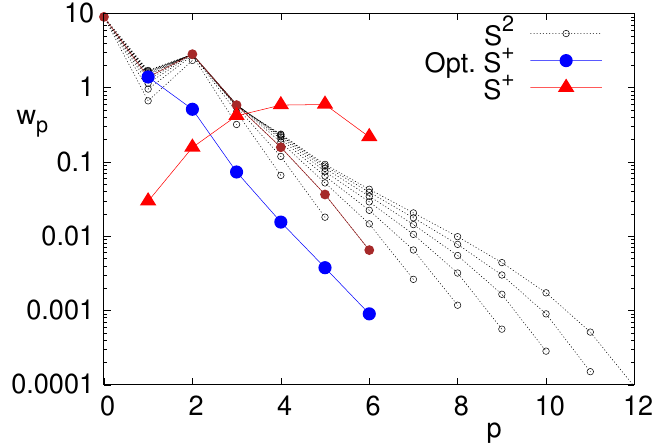}}
  \caption{{\bf Quasi few-body nature of SU(2) generators at $\Uq$ points.} The total weight $w_p$ of operators with $p$ non-identity Pauli operators is shown for the Casimir $S^2/L$ for $L=2,\ldots,12$ (black circles), showing quasi few-body nature of generators. Red triangles show a multi-body nature of $S^+/\sqrt{L}$ with an arbitrary gauge, while blue circles \new{demonstrate} quasi few-body nature for the optimal gauge \new{($L=6$)}. Parameters are $D=B=\Delta=1$, and $\tau_+\approx\frac{\pi}{4}1.464$, \new{for which Eq.(\ref{eq:Uq2}) with $k=1$ gives $q=\frac{\sqrt{6}-\sqrt{2}}{2}\approx 0.518$.}}
  \label{fig:S2}
\end{figure}

The important question is locality of $X$ and $Y$. In Fig.~\ref{fig:S2} we see that $S^+$ constructed according to the above prescription (red triangles) are non-local multi-body operators -- expanding them over products of Pauli matrices $S^+=\sum_{\boldsymbol{\alpha}} c_{\boldsymbol{\alpha}}\, \sigma_1^{\alpha_1}\cdots \sigma_L^{\alpha_L}$ we have terms with product of $L$ operators in a system of $L$ spins. However, it is important to note that we have a gauge freedom in constructing $S^+$; each eigenstate $\ket{m}$ is determined upto a multiplicative phase $\ee{\ii \varphi_m}$. It turns out that the choice of those phases can greatly influence the locality of $S^+$, while on the other hand locality of the Casimir operator $S^2=X^2+Y^2+Z^2$ does not depend on them. Exponential decay of $S^2$ (Fig.~\ref{fig:S2}) suggests that also $S^+$ can be chosen such that they involve only few-body terms. This is indeed the case. Writing $S^+$ with all possible phases $\varphi_m$ and numerically finding the optimal ones for which the weight of 1-body terms is the largest, we get $S^+$ for which the weight of many-body terms decays exponentially with the number of terms $p$ (blue points).

\begin{figure}[t]
  \centerline{\includegraphics[width=2.5in]{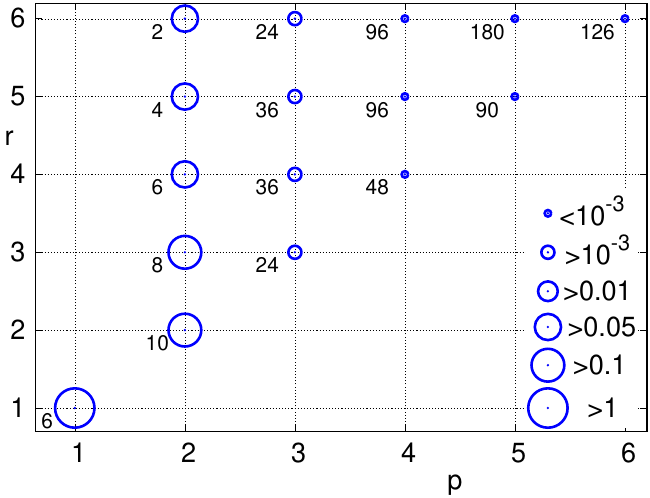}}
  \caption{{\bf Long-range nature of $S^+$.} Locality \new{properties} of $S^+$ for the optimal gauge choice and $L=6$ -- the same data as blue points in Fig.~\ref{fig:S2}. Circles indicate in log-scale the average weight $w$ of operators according to the number of non-identity terms $p$ and the range $r$ (numbers next to circles is the number of such terms).}
  \label{fig:SpXXZ}
\end{figure}

We can classify locality of $S^+$ according to the total weight $w =\sum |c_{\boldsymbol{\alpha}}|^2$ of operators with range $r$ and the number of non-identity terms $p$. For instance, the operator $\sx_1 \sz_4$ has $p=2$ and $r=4$. We see in Fig.~\ref{fig:SpXXZ} that the SU(2) generators constructed at $\Uq$ points and for the optimal choice of phases $\varphi_m$ are quasi few body -- weight decays exponentially with $p$ -- however, they contain operators with range $r \sim L$. The weight does not appreciably decay with $r$ for fixed $p$. Both these properties hold also for $\Uq$ generators (\ref{eq:genUq}). For instance, one-body operators ($p=r=1$) are $\sigma^+_l$ and have the average weight $1.4$, the nearest-neighbor 2-body terms \new{are} $\sigma^+_l \sz_{l\pm 1}$ with the average weight $0.14$, while 2-body range $r=6$ terms are $\sigma_1^+ \sz_L$ and $\sz_1 \sigma^+_L$ with the average weight $0.06$. The largest weight of any of 3-body term ($p=3$) is $0.03$, with the average of all 24 $r=p=3$ terms $0.007$.

\begin{figure*}[t!]
  \centerline{
    \includegraphics[width=\textwidth]{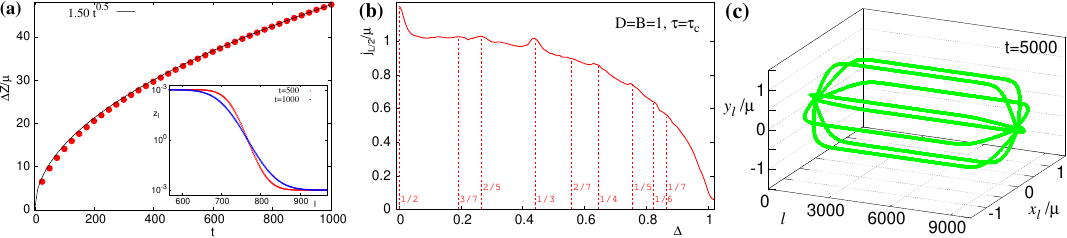}
  }
  \caption{{\bf Magnetization spreading.} (a) Diffusive transport at $\Uq$ point with $\tau_+\approx\frac{\pi}{4}\cdot 1.464$ (green triangle in Fig.~\ref{fig:all}(a)) despite the nonlocal SU(2) symmetry ($L=1536, \chi=576$). The inset shows magnetization profiles, the main plot transferred magnetization. (b) Fractal dependence on $\Delta$ at $\tauc\approx 0.60 \frac{\pi}{4}$ in the ballistic phase. All rational peaks $p/m$ (\ref{eq:peaks}) with $2 \le m \le 7$ are marked by vertical dashed lines ($L=4000$ and $t=1000$). (c) Magnetization in the $xy$ plane for a superdiffusive helix state at $t=5000$ and critical $\tauc$ and $\Delta=1$ (same data as in Fig.~\ref{fig:all}(c)), starting at $t=0$ with $\langle \tilde{\sigma}_l^{\rm x}\rangle=\mu$ and $\langle \tilde{\sigma}_l^{\rm y}\rangle=\langle \sz_l \rangle=0$. In all plots $B=D=1$.}
  \label{fig:all3}
\end{figure*}

\begin{flushleft}
  {\bf Numerical simulations}
\end{flushleft}
\vskip-3mm
\label{app:num}
Numerical simulation of unitary time evolution of density operators is performed by writing $\rho$ in a matrix product operator (MPO) form with matrices of maximal bond size $\chi$. Application of a single two-site gate is performed using standard time-evolved-block decimation (TEBD) techniques~\cite{vidal04}.

The initial state is a weakly polarized domain wall
\begin{equation}
  \rho(0) \propto \prod_{l=1}^{L/2} (\1+\mu \sz_l)\,\, \otimes\!\! \prod_{l=L/2+1}^{L} (\1-\mu \sz_l),
\end{equation}
with $\mu=10^{-3}$, \new{and we use OBC}. Such a state is numerically relatively easy to simulate, e.g., at the superdiffusive point we could get a relative precision of order $10^{-4}$ with $\chi=256$. On top of it, as explained in Ref.~\cite{kpz}, it can be used as a trick to calculate the equilibrium infinite-temperature autocorrelation function of $\sigma^{\rm z}$. If high precision is required one should stop simulation when the lightcone from the center touches a lightcone of small errors propagating from the edge. At the critical manifold this is a direct consequence of the screw SU(2) symmetry. The \new{open edge acts} as a defect due to a rotation caused by $D$ (\ref{eq:tSp}), nicely visible if one starts with a domain wall in the $xy$ plane, or, \new{even better, with a pure state fully polarized in the $\tilde{x}$} direction (Fig.~\ref{fig:all3}(c)). \new{Due to the SU(2) symmetry one would expect that such a state will be time independent, however, because one has a boundary violation of SU(2) symmetry (\ref{eq:comm}), a perturbation in the form of a helix (reflecting the screw-like nature of the symmetry) spreads away from boundaries. This can be seen Fig.~\ref{fig:all3}(c): for shown $D=1$ one has $2\vartheta=\pi/4$, i.e., $8$ sites are needed for a full turn of the helix. Because we plot all $10^4$ sites individual $l$ fuse into 8 distinct curves with consecutive $l$ lying at neighboring curves.}

\new{Note that while the exact integrability has been proved in Ref.~\cite{U1} only for PBC, for bulk physics like transport a potential boundary violation of integrability is immaterial. We expect that with an appropriate boundary gates one could achieve exact integrability also for OBC, similarly as has been done for the XXZ gate~\cite{vernier24}.}

We have checked transport at special parameter values in the diffusive phase I where one has $\Uq$ symmetry and therefore also SU(2) operators that commute with $U$. An example is show in Fig.~\ref{fig:all3}(a), where we can see that one gets diffusion. The non-locality of SU(2) generators is enough, despite being quasi few-body operators, to not affect transport of magnetization. We note that the required bond size needed to describe time evolution in those cases is rather large, for instance, $\chi$ more than $500$ at $t=1000$. In Fig.~\ref{fig:all3}(b) we show fractal dependence of the ballistic transport on $\Delta$.

\begin{flushleft}
  {\bf Current operator}
\end{flushleft}
\vskip-3mm
\label{app:cur}

The spin current operator is defined via a discrete-time continuity equation
\begin{equation}
  U^\dagger Z_{[k,l]} U - Z_{[k,l]}= j_{k-1}-j_l,
\end{equation}
where $j_k$ is the local current operator between sites $k$ and $k+1$. For the brickwall circuit $U$ is invariant under translations by $2$ sites and we will have different current operator on even and odd sites. Specifically, the current on even sites, i.e., on bonds between the legs of the 1st layer gates (Fig.~\ref{fig:phases}), can be identified from $U^\dagger (\sz_3+\sz_4) U- (\sz_3+\sz_4)=j_2-j_4$, and is explicitly
\begin{eqnarray}
  j_{2}&=&{\cal A}(\sx_2 \sy_3-\sy_2 \sx_3)+{\cal F}(\sz_2-\sz_3)+{\cal C}(\sx_2\sx_3+\sy_2\sy_3) \nonumber \\
  \quad && {\cal A}=\frac{\sin{(4\tau\Jeff)}}{2\Jeff}+\frac{BD\sin^2{(2\tau\Jeff)}}{\Jeff^2},\nonumber \\
  && {\cal F}=\frac{(1+D^2)\sin^2{(2\tau\Jeff)}}{\Jeff^2},\nonumber \\
  \qquad && {\cal C}=\frac{B\sin^2{(2\tau\Jeff)}}{\Jeff^2}-\frac{D\sin{(4\tau\Jeff)}}{2\Jeff}.
\label{eq:jdef}
\end{eqnarray}
The current on odd sites is more complicated, in general a $4-$site operator that can be calculated similarly.

\vskip-2mm
\begin{flushleft}
  {\bf Symmetries}
\end{flushleft}
\vskip-2mm
\label{app:sym}

Here we list some other symmetries of the model. Rotation by $W$ (\ref{eq:W}), which is a Hamiltonian version of the twist transformation known from the \new{$R$-matrix of integrable} 6-vertex model~\cite{reshetikhin90,kulish09,U1}, can transform $D$ to zero, \new{$WU'W^\dagger=U(\tau,\Delta,B,D,m)$, with parameters of $U'$ being} $D'=0$ and $\tau'=\tau\sqrt{1+D^2}$, $\Delta'=\Delta/\sqrt{1+D^2}$, $B'=B/\sqrt{1+D^2}$, $m'=m/\sqrt{1+D^2}$. Rotation by $\vartheta=\pi/2$, $W_{\pi/2}=(-\ii)^{L(L+1)/2}\prod_{l} \sz_{2l-1}$, $U'=W_{\pi/2}^\dagger U W_{\pi/2}$, instead flips all parameters except $D$, $\tau'=-\tau$, $\Delta'=-\Delta$, $D'=D$, and $B'=-B$. Particle-hole transformation $P=\prod_l \sx_l$, $U'=P^\dagger U P$, changes the sign of two chiral terms, $\tau'=\tau$, $\Delta'=\Delta$, $D'=-D$ and $B'=-B$. Spatial reflection $R$ that changes site $l$ to $L+1-l$, $U'=R^\dagger U R$, does the same as $P$. This means that $U$ is invariant under the combined $\mathbbm{Z}_2$ symmetry $RP$.\\

{\em \bf Acknowledgements}\\
The author would like to thank Enej Ilievski, Toma\v{z} Prosen and Sebastian Diehl for discussions, and acknowledge Grants No.~J1-4385 and No.~P1-0402 from Slovenian Research Agency (ARIS). Hospitality of the Kavli Institute for Theoretical Physics (KITP) and support by grant NSF PHY-2309135 is also appreciated.\\

\end{document}